\documentclass[letterpaper]{article} 
\usepackage{aaai22}  
\usepackage{times}  
\usepackage{helvet}  
\usepackage{courier}  
\usepackage[hyphens]{url}  
\usepackage{graphicx} 
\usepackage{natbib}  
\usepackage{caption} 
\DeclareCaptionStyle{ruled}{labelfont=normalfont,labelsep=colon,strut=off} 
\usepackage{algorithm}
\usepackage{algorithmic}

\usepackage{makecell}
\usepackage{color}
\usepackage{graphicx}

\title{Adversarial Zoom Lens: A Camera-based Physical Attack to DNNs}

\author {
    Chengyin Hu \textsuperscript{\rm 1},
    Weiwen Shi \textsuperscript{\rm 1}
}
\affiliations {
    \textsuperscript{\rm 1} University of Electronic Science and Technology of China\\
    cyhuuestc@gmail.com, Weiwen\_shi@foxmail.com
}

\usepackage{bibentry}

\begin{document}

\maketitle

\begin{abstract}
Although deep neural networks (DNNs) are known to be fragile, there are few studies on the effects of camera-based physical attacks on DNNs' performance. In this paper, we demonstrate an effective physical-world attack called Adversarial Zoom Lens (\textbf{AdvZL}), which manipulates a zoom lens to zoom in and out of objects to generate adversarial samples, fooling DNNs without modifying the target objects. We use zoom-in images to verify the adversarial effect of AdvZL in the digital environment. Then, we use a zoom lens to generate adversarial samples, verifying the adversarial performance of AdvZL in the physical environment. On the other hand, we discuss some phenomena generated by AdvZL. Finally, we look into the possible threat of the proposed approach to future autonomous driving and variant attack approaches similar to the proposed attack.
\end{abstract}

\section{Introduction}
\label{sec: Introduction}
On the night of March 9, 2018, an unusual car accident occurred when a woman crossing the street was hit and killed by an SUV. What's unusual about the accident is that the SUV was operating on Autopilot, a system developed by Uber. The accident is believed to be the world's first fatal driverless crash. It can be seen that advanced DNNs make mistakes even if there is no external interference. If further attackers maliciously interfere with the classification system, it could lead to more accidents.

At present, the attack and defense technologies driven by adversarial attack are hot topics for many researchers. Most researchers focus on adversarial attacks in the digital environment \cite{ref18,ref19,ref20,ref21}, which fool advanced DNNs by adding imperceptible perturbations to clean images. However, in physical scenes, images taken by the camera are input into the classifier for classification, attackers cannot directly modify the input images. Many researchers gradually devote themselves to the study of physical attacks \cite{ref22,ref23,ref24}. The physical perturbation is designed to be much larger than digital one, so that it could be sensed by camera. Some physical attacks \cite{ref24,ref25}, use stickers and graffiti as perturbations to perform attacks while maintaining the semantic information of the target objects, which are perceptible to human observers. In addition, some researchers use light \cite{ref35,ref36} to carry out adversarial attacks, which to a certain extent achieves excellent concealment. Different from the existing methods, our proposed method is camera-based and fools the current advanced DNNs without modifying the target objects.

\begin{figure}
\centering
\includegraphics[width=1\columnwidth]{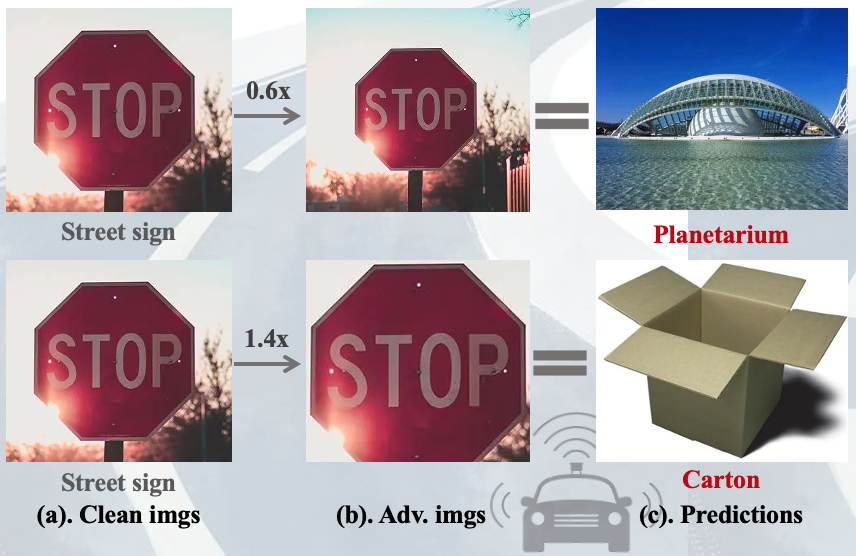} 
\caption{An example. A self-driving car equipped with a zoom lens is driving on the road. When passing the Stop sign, the zoom lens zooms in and out of the image, leading the classifier to misclassify the Street sign as Planetarium and Carton, etc.}.
\label{figure1}
\end{figure}

The difficulties of physical attacks include: (1) It’s difficult for camera to capture the printed pixel-level digital perturbations; (2) It is difficult to print perfectly adversarial graffiti (there exists printing loss); (3) Physical attacks are difficult to achieve both robustness and concealment.

Based on the above challenges, we propose a camera-based physical adversarial attack called Adversarial Zoom Lens (AdvZL). Different from the existing methods, we manipulate the zoom lens to zoom in and out on the target objects, realizing effective physical attacks without modifying the target objects. AdvZL fundamentally solves the above difficulties of physical attacks by not adding perturbations to target objects. we regard the physical attack shown in Figure \ref{figure1} as a novel adversarial attack, which is non-negligible but not yet exploited. An automatic zoom lens mounted on a self-driving car camera that zooms in and out on road signs, the self-driving car fails to recognize them correctly.

We perform comprehensive experiments in both digital and physical environments to verify the effectiveness of AdvZL. In a digital environment, we construct datasets of 1000 images from ImageNet \cite{ref45} that can be correctly classified by DNNs for test. Experimental results verify the effectiveness of AdvZL against advanced DNNs. In the physical environment, we use the zoom lens to zoom in and out target objects. Experimental results show that AdvZL achieves a 100\% success rate of physical adversarial attack within a certain range of distance and angle. Our main contributions are summarized as follows:

\begin{itemize}
\item We propose a camera-based physical attack, AdvZL, which manipulates the zoom lens to carry out physical attacks without modifying the target objects. The proposed method is rather simple to deploy, manipulating the zoom lens to conduct adversarial attacks (See Introduction).
\item We summarize the existing physical attacks (See Related work), carry out strict experimental design and comprehensive experimental test. Experimental results verify the effectiveness of AdvZL in the both digital and physical environments (See Approach, Evaluation).
\item We explore some AdvZL-based phenomena, which will help scholars reseach the mechanism of advanced DNNs and study defense strategies against AdvZL (See Discussion). At the same time, we look into some new ideas for camera-based physical attacks (See Conclusion).
\end{itemize}

\section{Related work}
\subsection{Digital attacks}
Adversarial attack was first proposed by Szegedy et al. \cite{ref1}, after which more and more adversarial attacks were proposed successively \cite{ref16,ref17,ref20,ref21}. 

At present, most digital attacks ensure the perturbations are imperceptible to human observers by limiting them in a norm-ball. In general, ${L}_{2}$ and ${L}_{\infty}$ are the most commonly used norms \cite{ref2,ref3,ref4,ref5,ref6}, these methods effectively attack advanced DNNs in the digital environment while ensuring the perturbations are imperceptible to human observers. Some other works have modified other attributes of the clean sample to generate adversarial samples, such as color \cite{ref7,ref8,ref9}, texture and camouflage \cite{ref10,ref11,ref12,ref13}, which are usually perceptible to the naked eye. In addition, there are also some works to generate adversarial samples by modifying the physical parameters of the clean images \cite{ref14,ref15}, which retain the key components of images and carry out digital attacks. Different from digital attacks that modify the pixels of clean images in the digital environment, physical attacks cannot directly modify the input images.

\subsection{Physical attacks}
Physical attack was first proposed by Alexey Kurakin et al. \cite{ref22}. After this work, many physical attacks were proposed successively \cite{ref24,ref28,ref29,ref30,ref31}.

\textbf{Traditional street sign attacks.} Ivan Evtimov et al. \cite{ref24} proposed a general physical adversarial attack, called RP2, that achieved a robust physical attack against road sign classifiers. However, RP2 is susceptible to environmental interference at large distances and angles. Chen et al. \cite{ref23} proposed ShapeShifter, which successfully fooled classifiers by generating stop signs of reverse interference. Huang et al. \cite{ref27} improved ShapeShifter by adding Gaussian white noise to ShapeShifter's optimization function. The experiment proved that the improved ShapeShifter could successfully and effectively attack Chinese and English stop signs, and overcome the shortcoming of ShapeShifter's high requirements for photographic equipment. However, ShapeShifter and the improved ShapeShifter have a defect, disturbance covers almost the whole road sign, failed to achieve concealment. Eykholt et al. \cite{ref26} implemented a disappear attack, in which poster and stickers were stickled on the surface of road signs to fool the target detector and realize transferable adversarial perturbations. Similarly, the perturbations cover a large area, which is too conspicuous. Duan et al. \cite{ref25} proposed AdvCam, which camouflages physical perturbations in a natural style and achieves better camouflaging effect while fooling classifiers. AdvCam has better concealment than the above methods, but it needs to manually select the attack area and target. On the whole, traditional road sign attacks have something in common, which is, physical perturbations are printed or pasted onto the road signs. These methods have major drawbacks. Adding physical perturbations is manual work, takes a lot of time, at the same time, there exists printing errors.

\begin{figure*}
\centering
\includegraphics[width=1\linewidth]{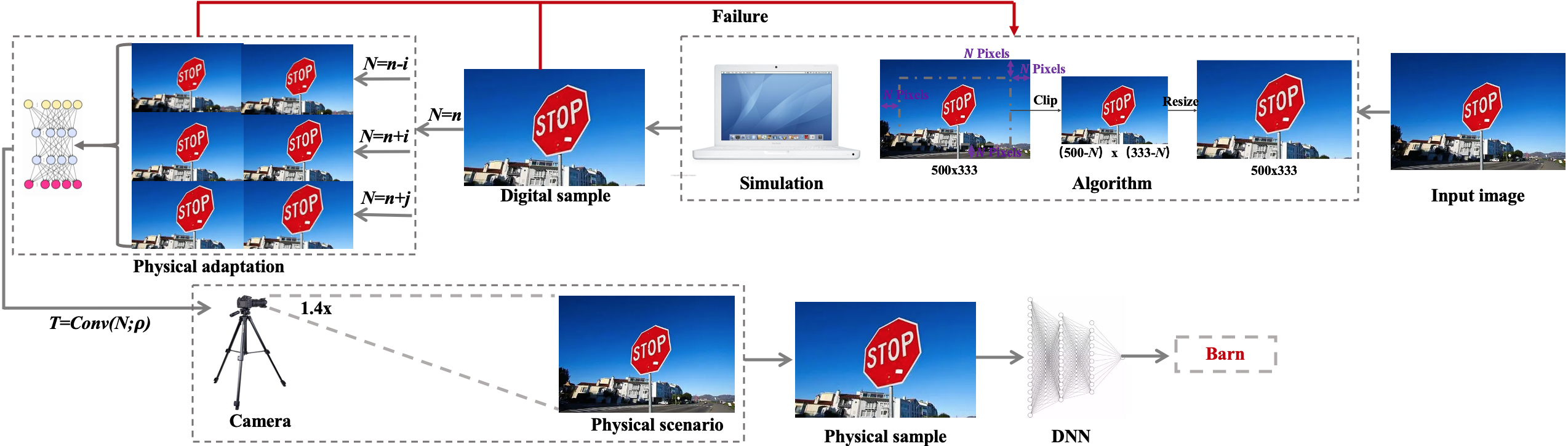} 
\caption{Generating an adversarial sample.}.
\label{figure2}
\end{figure*}

\textbf{Light-based attacks.} Light-based physical attack is a crafty solution to above problem. Nguyen et al. \cite{ref33} studied the threat of light projection to face recognition system, and implemented white-box and black-box attacks against face recognition system by projecting a well-designed adversarial light projection onto human face. But it's complex for deployment. Shen et al. \cite{ref32} proposed VLA, which is based on visible light and uses a carefully designed light beam to fool the face recognition system, enabling targeted or untargeted attacks. On the contrary, Zhou et al. \cite{ref34} generated adversarial samples based on invisible infrared light, and were the first to interpret the threat of infrared adversarial samples to face recognition system. These attacks achieve better concealment, but modify the target objects. Duan et al. \cite{ref35} proposed AdvLB, which achieves efficient and covert physical attacks by manipulating the physical parameters of the laser beam. Gnanasambandam et al. \cite{ref36} proposed OPAD, which realized effective optical adversarial attacks against 2D and 3D objects. AdvLB and OPAD, however, can only perform attacks in weak-light conditions. Zhong et al. \cite{ref37} studied a new type of optical adversarial sample, using a very common natural phenomenon, shadow, generate adversarial sample, to achieve a natural and hidden black box adversarial attack. But it is difficult to work in complex physical scenes. Overall, light-based methods allow for more efficient physical adversarial attacks and better invisibility. However, they require a variety of colors of light as physical perturbations that are perceptible to human observers. In addition, light-based physical attacks tend to paralyze during the daytime.

\textbf{Camera-based attacks.} To avoid modifying the target objects, Li et al. \cite{ref38} study the physical operation of the camera itself, through an iterative update against perturbations, then elaborate translucent stickers affixed to the camera lens, shooting target objects to generate adversarial samples, this method is to inject perturbations into the optical path between the camera and the object, which has excellent concealment. However, it is difficult to adapt to complex real scenes due to the complicated deployment. Our proposed method (AdvZL) fools the advanced DNNs by manipulating the zoom lens to zoom in and out of the target objects and generate adversarial samples.

\section{Approach}
\subsection{Adversarial sample}
Reviewing the definition of adversarial sample ${X}_{adv}$, given an input image $X$, ground truth label $Y$, a DNN classifier $f$, $f(X)$ represents the predicted label, the classifier $f$ associates with a confidence score ${f}_{Y}(X)$ to class $Y$. Generating adversarial samples satisfies two properties : (1) $f({X}_{adv} \neq f(X) = Y$; (2) $\parallel {X}_{adv} - X \parallel < \epsilon$. The first requires that ${X}_{adv}$ successfully fool DNN classifier $f$, and the second requires that the adversarial perturbations be small enough to be imperceptible to the human observers.

Different from most of the existing physical attacks, in this paper, we generate adversarial sample by zooming in or out the target objects, which fools the advanced DNNs classifier without adding perturbations to the target objects.

\subsection{Zoom lens definition}
\textbf{Zoom in.} In the digital environment, as shown in Figure \ref{figure2}, $N$ represents the pixel value to be clipped, the larger $N$ is, the greater zoom is. In the physical environment, in order to keep the physical samples and digital samples consistent, we define an function to convert the pixel value $N$ to the camera zoom factor $T$, which can be expressed as:

\begin{equation}
    \label{Formula 1}
    T = conv(N;\rho) = {[N/\rho]}_{1}
\end{equation}

Where, ${[/]}_{1}$ indicates that the division takes one decimal place and the second decimal place is rounded. $\rho$ indicates that for every 0.1x zoom of the camera, the number of pixels needed to be cropped in the digital environment is $\rho$ pixel. Through Function \ref{Formula 1}, we realize the conversion between digital samples and physical samples.

\textbf{Zoom out.} In the digital environment, we cannot realize the zoom out attack. Therefore, we perform the zoom out attack experiment in the physical environment. Here, we use camera to take the zoom-out photos, such as the camera's 0.1x focal length. The image is then used for step-to-step zoom in attacks. The adversarial samples below 1.0x focal length is the adversarial samples of the zoom out attack.

\begin{figure*}
\centering
\includegraphics[width=1\linewidth]{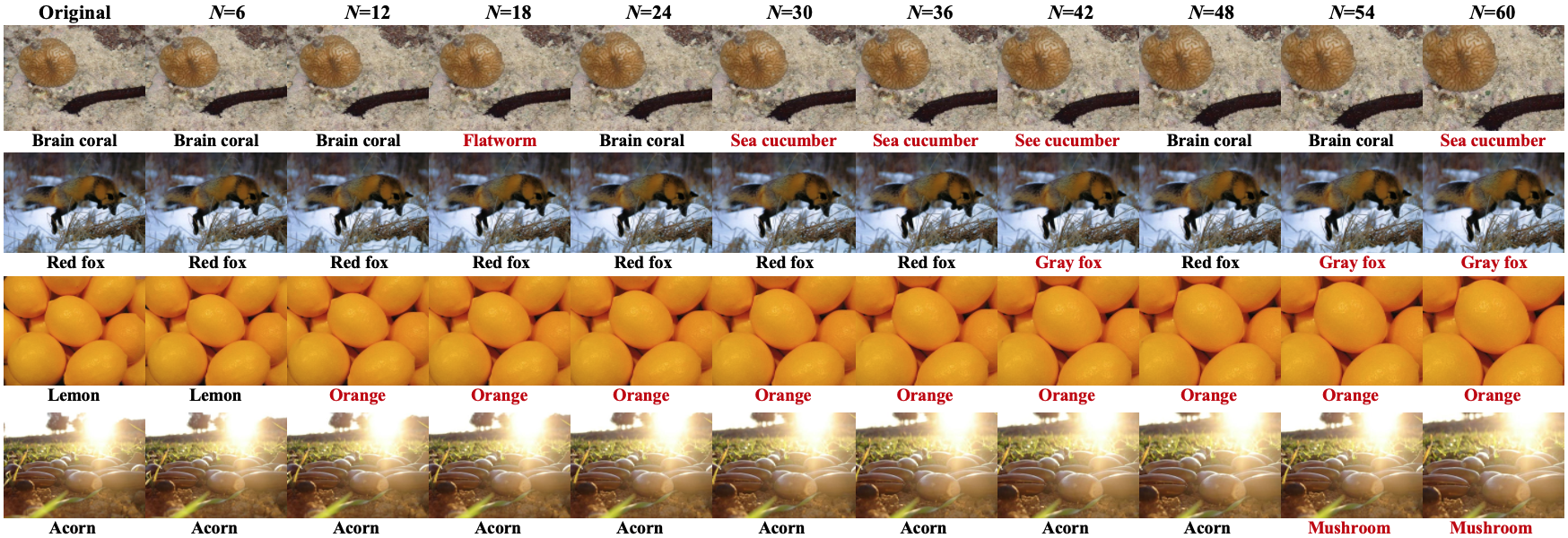} 
\caption{Digital adversarial samples.}.
\label{figure3}
\end{figure*}

\subsection{Zoom lens adversarial attack}
In the digital environment, we perform zoom in attacks on 1000 images selected from ImageNet [45] that could be correctly classified by each advanced DNN, verifying the adversarial effect of the zoom-in image to DNNs. In the physical environment, we manipulate the zoom lens to zoom in and out the target objects, which verifies the feasibility of AdvZL in the physical environment. In the digital environment, generating an adversarial sample is expressed as follows:

\begin{equation}
    \label{Formula 2}
    {X}_{adv} = {ZoomIn}_{N}(X) \quad s.t. \quad N \in [0, \omega]
\end{equation}

Where $N$ represents the magnification degree of the image and $\omega$ represents the threshold of the magnification degree. In the physical environment, zoom lens is used to zoom in and out the target objects to generate adversarial sample, generating an adversarial sample can be written as follows:

\begin{equation}
    \label{Formula 3}
    {X}_{adv} = {Zoom}_{T}(X) \quad s.t. \quad
    T \in [{\Gamma}_{min},{\Gamma}_{max}]
\end{equation}

Here, $T$ is a multiple of zooming in or out (e.g., ${Zoom}_{0.8\times}(X)$ indicates that the image X is zoomed out to 0.8 times, and ${Zoom}_{1.2\times}(X)$ indicates that the image is zoomed in to 1.2 times), ${\Gamma}_{min}$ and ${\Gamma}_{max}$ indicate the threshold of $T$.

\textbf{Physical Adaptation.} To solve the experimental loss from digital sample to physical adversarial sample, we define an $Adjust$ function, the operations include increase and decrease the pixel value of $N$, which is expressed as:

\begin{equation}
    \label{Formula 4}
    Adjust({X}_{adv}; N)
\end{equation}

In this experiment, we consider a practical situation: the attacker cannot obtain the knowledge of the model, but only the confidence score ${f}_{Y}(X)$ with given input image $X$ on ground truth label $Y$. In our proposed method, we use confidence score as the adversarial loss. Thus, the objective is formalized as minimizing the confidence score on the ground truth label $Y$, which can be formulated as follows:

\begin{equation}
    \label{Formula 5}
    \mathop{\arg\min}_{N}{f}_{Y}({ZoomIn}_{N}(X)) \quad s.t. \quad N \in [0,\omega]
\end{equation}

\textbf{Algorithm.} 
As shown in Algorithm \ref{algorithm1}. The proposed AdvZL takes clean sample $X$, ground truth label $Y$, threshold ${\Gamma}_{min}$ and ${\Gamma}_{max}$, classifier $f$ as input. Then, zooming in and out the clean sample $X$ to different degrees, the adversarial sample with the smallest confidence score of classifier $f$ on ground truth label $Y$ is taken as the most adversarial one. The algorithm finally returns an adversarial sample fooled advanced DNNs, which is used to perform subsequent physical attacks. Here, the value range of $T$ is generally from 0.5 to 5.4, with an interval of 0.1. Therefore, ${\Gamma}_{min}$ and ${\Gamma}_{max}$ should satisfy: ${\Gamma}_{min} \geq 0.5$, ${\Gamma}_{max} \leq 5.4$.

\begin{algorithm}
	\renewcommand{\algorithmicrequire}{\textbf{Input:}}
	\renewcommand{\algorithmicensure}{\textbf{Output:}}
	\caption{Pseudocode of AdvZL}
	\label{algorithm1}
	\begin{algorithmic}[1]
	
		\REQUIRE Input $X$, Label $Y$, ${\Gamma}_{min}$, ${\Gamma}_{max}$, Classifier $f$;
		\ENSURE Adversarial sample ${X}_{adv}^{\star}$;

		\STATE \textbf{Initialization} ${X}_{adv}^{\star} = X$, ${Score}^{\star} = {f}_{Y}(X)$;

		\FOR{$T$ in range (${\Gamma}_{min}$, ${\Gamma}_{max}$)}
		    \STATE ${X}_{adv} = {Zoom}_{T}(X)$;
		    \STATE $Score = {f}_{Y}({X}_{adv})$;
		    
		    \IF{${Score}^{\star}<Score$}
		        \STATE ${Score}^{\star}=Score$;
		        \STATE ${X}_{adv}^{\star} = {X}_{adv}$
		    \ENDIF
		\ENDFOR
		
		\IF{$ argmaxf({X}_{adv}^{\star}) \neq argmaxf(X)$}
		    \IF{$Adjust({X}_{adv}^{\star}; N)$}
		        \STATE \textbf{return} ${X}_{adv}^{\star}$;
		    \ENDIF
		\ENDIF

	\end{algorithmic}  
\end{algorithm}

\begin{table*}[htbp]
    \centering
    \caption{\label{Table 1}Attack success rate (ASR) in the digital environment.}
    \begin{tabular}{ccccccc}
    \hline
    $f$ & DenseNet & ResNet50 & VGG19 & GoogleNet & MobileNet v2 & AlexNet\\
    \hline
    ASR(\%) & 36.0 & 33.7 & 40.5 & 40.6 & 41.7 & 51.0\\
    \hline
    \end{tabular}
\end{table*}

\section{Evaluation}
\subsection{Experimental setting}
We conduct comprehensive experiment to verify the effectiveness of AdvZL in both digital and physical environments. In the digital environment, we use the advanced DNNs classifiers for experiments. As with the approach in AdvLB \cite{ref35}, we use 1000 images from ImageNet \cite{ref45} that could be correctly classified by advanced DNNs as the dataset, although we cannot zoom out the image in the digital environment, we verify the adversarial effect of zoom-out sample in the physical setting. In the physical environment, we use ResNet50 as a target model to conduct experiments, with common objects and road signs as experimental objects. We use a mobile phone camera (iPhone6s) as a zoom lens for all of our experiments. It has been verified that the effectiveness of AdvZL is not affected by using a normal camera or other mobile phone model for our experiments.

\subsection{Evaluation of AdvZL}
Digital test. We test the effectiveness of AdvZL on 6 subsets that randomly selected from ImageNet \cite{ref45}, each subset containing 1000 clean samples that could be correctly classified by the corresponding advanced DNN model.  Table 1 shows the attack success rate of advanced DNNs on the corresponding dataset.

As can be seen from the experimental results in Table \ref{Table 1}, AdvZL achieves an attack success rate of over 30\% against advanced DNNs \cite{ref39,ref40,ref41,ref42,ref43,ref44} without modifying the semantic information of the clean samples, which reduce the Top-1 classification accuracy from 100\% to less than 70\%. Figure \ref{figure3} shows the adversarial samples in the digital environment (ResNet50 is used here). It shows that with the image is enlarged, the advanced DNN model misclassifies the images. For example, Brain coral is misclassified as Flatworm, Sea cucumber, etc. In general, without modifying the semantic information of the image, the larger magnification of the image, the stronger adversarial effect to DNNs. Figure \ref{figure3} also reveals a common phenomenon, advanced DNNs are trained in a dataset in which the images are taken at a specific range of distance. Changing the distance to take a picture is equivalent to zooming in and out of the picture, which means that when zooming in and out of the picture, the classifier makes a wrong classification judgment.


\begin{figure}[H]
\centering
\includegraphics[width=0.8\columnwidth]{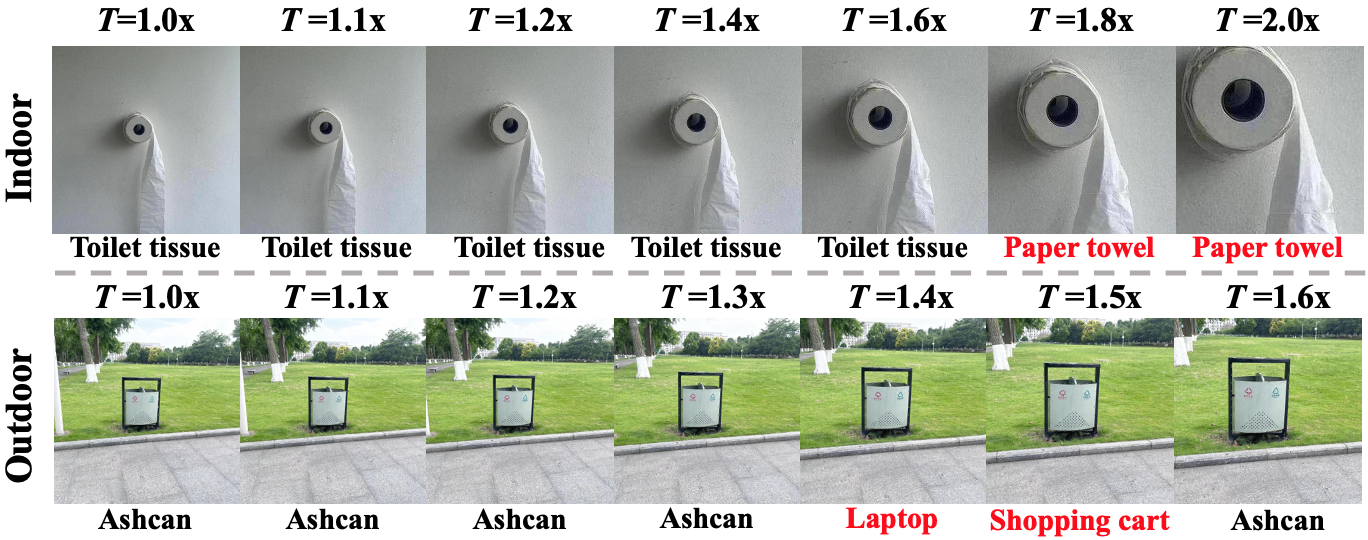} 
\caption{Indoor test and outdoor test.}.
\label{figure4}
\end{figure}

\textbf{Physical test.} We use ResNet50 to perform all the physical tests. In order to execute untargeted attack in the physical environment, it is necessary to eliminate the influence of experimental noise, we design indoor test and outdoor test respectively. In which, physical attacks are not affected by the environmental noise in the indoor test, well, much of the environmental noise may play the role of an adversarial perturbation in the outdoor test. Through our comprehensive experiments, we verified the effectiveness of AdvZL in the both indoor and outdoor environment. In our experiment, Toilet tissue and Ashcan, which are common in daily life, are selected as the testing object. Figure \ref{figure4} shows the schematic diagram of the adversarial sample after Zooming in the image in the physical environment.

\begin{figure}[H]
\centering
\setlength{\abovedisplayskip}{-0.9cm}
\includegraphics[width=1\columnwidth]{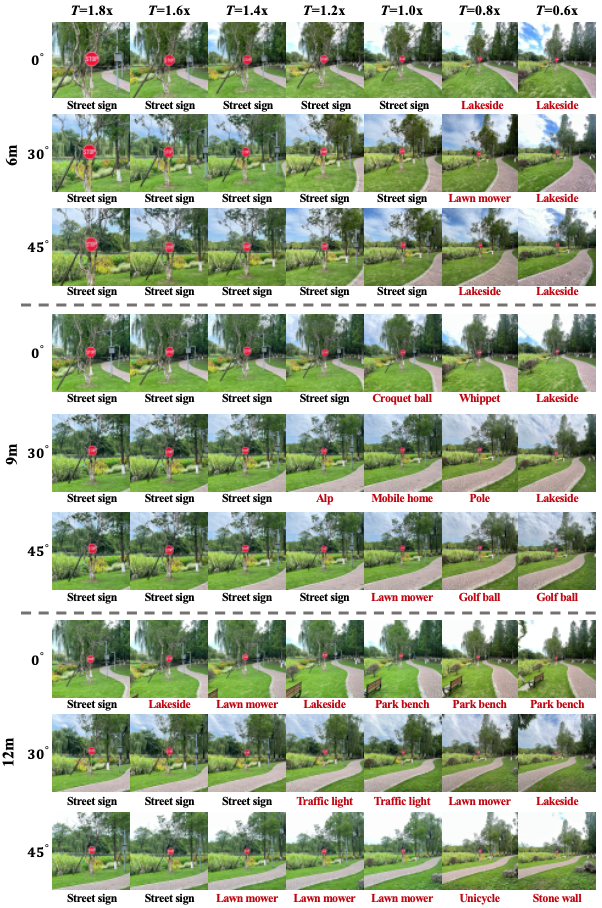} 
\caption{Adversarial samples at different distances and angles.}.
\label{figure5}
\end{figure}

As can be seen from Figure \ref{figure4}, adversarial samples deceive the advanced DNN model as $T$ increases to the threshold. For example, when Ashcan is magnified to $T=1.4\times$ and $T=1.5\times$ of the focal length of the mobile phone camera, the advanced DNN model misclassifies them as Laptop and Shopping cart respectively. Here, $1.4\times$ represents the zoom multiple of the camera. To get close to the real scenario, we design a comprehensive attack experiment for the outdoor test, where the attacking target object is the common Stop sign. We carry out physical attacks at different distance and angle. Figure \ref{figure5} shows the schematic diagram of partial adversarial samples at different distances and angles, among which, we test adversarial samples at distances of 6m, 9m, 12m and angles of ${0}^{\circ}$, ${30}^{\circ}$ and ${45}^{\circ}$ respectively.

As can be seen from Figure \ref{figure5}, at different distances, we manipulate the zoom lens to zoom out the image, adversarial samples from different angles fool the target model. Obviously, even though the images contain environmental noise such as Lakeside and Pole, human observers tend to ignore the environmental noise and more inclined to consider them as Street sign. When the distances are 9 meters and 12 meters, it shows that the background environmental noise increases, which makes the physical samples more adversarial. A slight manipulation of the zoom lens generates an adversarial sample to fool advanced DNN model. In the real scenario, since it is difficult to capture the road signs with a long distance, the lens of the autonomous vehicle is properly adjusted during production. Therefore, the effectiveness of AdvZL is not affected even at a distance of 9 meters and 12 meters.


\section{Discussion}
Here, we show some phenomena generated by AdvZL in the both digital and physical environment. All of the experiments in this section use ResNet50 as the target model.

\subsection{Discontinuous misclassification}
\label{section5_1}
\begin{figure}[H]
\centering
\includegraphics[width=1\columnwidth]{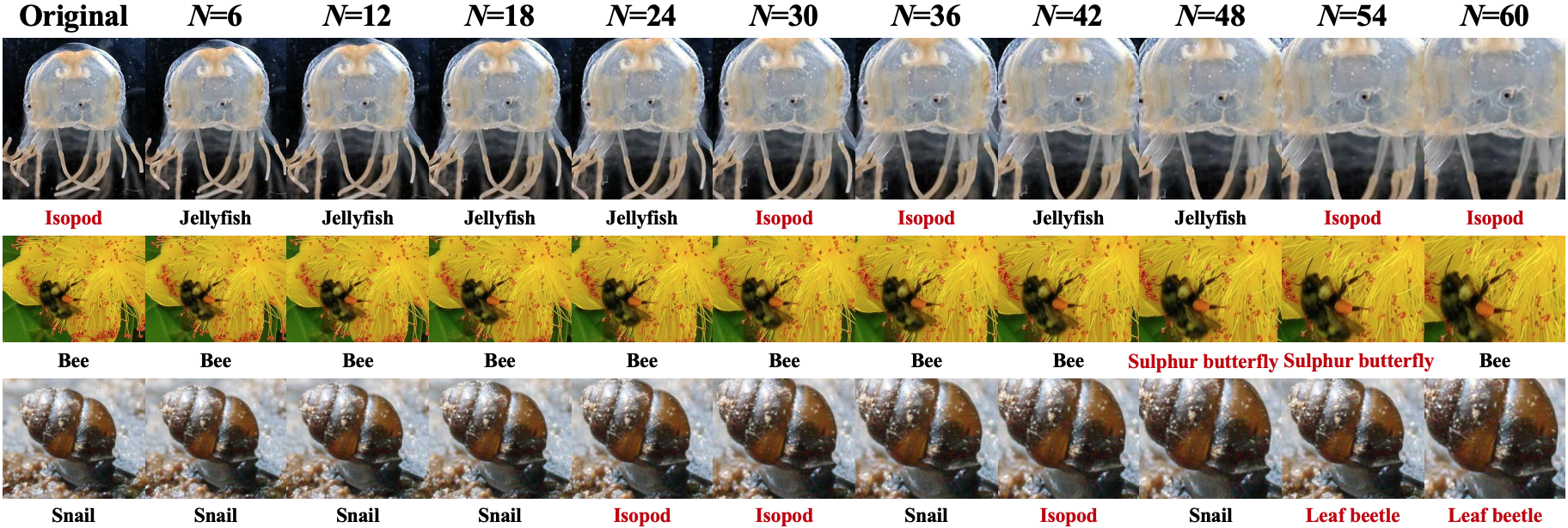} 
\caption{Discontinuous misclassification in the digital environment.}.
\label{figure6}
\end{figure}

As can be seen from the experimental results in Figure \ref{figure3}, we zoom in the clean samples with equal proportions, discontinuous classification errors will occur instead of continuous ones. In view of this phenomenon, we analyze the experimental data. It can be seen from Figure \ref{figure6} that :(1) Some clean samples could not have been classified correctly by the model, but corectly classified after zooming in. (2) Classification errors may occur in some images at a specific magnification. As for the Bee in Figure 6, when $N=48$ and $N=54$, it occurs classification errors. (3) Discontinuous misclassification. As shown in Figure \ref{figure6}, Snail is classified as Isopod when $N=30$, Snail when $N=36$, Isopod when $N=42$, Snail when $N=48$, Leaf beetle when $N=54$.

In general, the phenomenon in Figure \ref{figure6} reflects, to some extent, that the decision boundary between various categories of classifiers is relatively dense, which leads to the advanced DNN model is prone to errors under slight confrontation. Therefore, the security implications of advanced DNNs should be of great concern.

Similarly, discontinuous misclassification may also occur in the physical environment. The data in table 2 shows the classification results of ResNet50 for adversarial samples at different distances and angles. 919 represents the digital label of Street sign. Here, the samples are common road signs in daily life. It can be seen that discontinuous misclassification of ResNet50 occurs when the distance and angle are (9m, ${45}^{\circ}$), (12m, ${0}^{\circ}$), (12m, ${45}^{\circ}$) respectively.

\begin{table*}[htbp]
    \centering
    \caption{\label{Table 2}Discontinuous misclassification in the physical environment.}
    \begin{tabular}{cccccccccc}
    \hline
    
    \quad & $(6m, {0}^{\circ})$ & $(6m, {30}^{\circ})$ & $(6m, {45}^{\circ})$ & $(9m, {0}^{\circ})$ & $(9m, {30}^{\circ})$ & $(9m, {45}^{\circ})$ & $(12m, {0}^{\circ})$ & $(12m, {30}^{\circ})$ & $(12m, {45}^{\circ})$\\
    \hline
    
    $0.5\times$ & \textcolor{red}{975} & \textcolor{red}{975} & \textcolor{red}{970} & \textcolor{red}{975} & \textcolor{red}{975} & \textcolor{red}{975} & \textcolor{red}{703} & \textcolor{red}{975} & \textcolor{red}{825} \\
    \hline
    
    $0.6\times$ & \textcolor{red}{975} &\textcolor{red}{975} & \textcolor{red}{975} & \textcolor{red}{975} & \textcolor{red}{975} & \textcolor{red}{574} & \textcolor{red}{703} & \textcolor{red}{975} & \textcolor{red}{825} \\
    \hline
    
    $0.7\times$ & \textcolor{red}{975} & \textcolor{red}{975} & \textcolor{red}{975} & \textcolor{red}{671} & \textcolor{red}{621} & \textcolor{red}{621} & \textcolor{red}{975} & \textcolor{red}{975} & \textcolor{red}{621} \\
    \hline
    
    $0.8\times$ & \textcolor{red}{975} & \textcolor{red}{621} & \textcolor{red}{975} & \textcolor{red}{172} & \textcolor{red}{733} & \textcolor{red}{574} & \textcolor{red}{703} & \textcolor{red}{621} & \textcolor{red}{880} \\
    \hline
    
    $0.9\times$ & 919 & \textcolor{red}{522} & \textcolor{red}{920} & \textcolor{red}{417} & \textcolor{red}{920} & 919 & \textcolor{red}{703} & \textcolor{red}{920} & \textcolor{red}{975} \\
    \hline
    
    $1.0\times$ & 919 & 919 & 919 & \textcolor{red}{522} & \textcolor{red}{660} & \textcolor{red}{621} & \textcolor{red}{703} & \textcolor{red}{920} & \textcolor{red}{621} \\
    \hline
    
    $1.1\times$ & 919 & 919 & 919 & \textcolor{red}{574} & \textcolor{red}{863} & \textcolor{red}{621} & \textcolor{red}{843} & \textcolor{red}{920} & \textcolor{red}{621} \\
    \hline
    
    $1.2\times$ & 919 & 919 & 919 & 919 & \textcolor{red}{970} & 919 & \textcolor{red}{975} & \textcolor{red}{920} & \textcolor{red}{621} \\
    \hline
    
    $1.3\times$ & 919 & 919 & 919 & 919 & 919 & 919 & \textcolor{red}{975} & 919 & \textcolor{red}{621} \\
    \hline
    
    $1.4\times$ & 919 & 919 & 919 & 919 & 919 & 919 & \textcolor{red}{621} & 919 & \textcolor{red}{621} \\
    \hline
    
    $1.5\times$ & 919 & 919 & 919 & 919 & 919 & 919 & 919 & 919 & 919 \\
    \hline
    
    $1.6\times$ & 919 & 919 & 919 & 919 & 919 & 919 & \textcolor{red}{975} & 919 & 919 \\
    \hline
    
    $1.7\times$ & 919 & 919 & 919 & 919 & 919 & 919 & 919 & 919 & 919 \\
    \hline
    
    $1.8\times$ & 919 & 919 & 919 & 919 & 919 & 919 & 919 & 919 & 919 \\
    \hline
    
    $1.9\times$ & 919 & 919 & 919 & 919 & 919 & 919 & 919 & 919 & \textcolor{red}{621} \\
    \hline
    
    $2.0\times$ & 919 & 919 & 919 & 919 & 919 & 919 & 919 & 919 & 919 \\
    \hline
    
    $...$ & ... & ... & ... & ... & ... & ... & ... & ... & ... \\
    \hline
    
    $2.5\times$ & 919 & 919 & 919 & 919 & 919 & 919 & 919 & 919 & 919 \\
    \hline
    \end{tabular}
\end{table*}

\subsection{Model attention}
\begin{figure}[H]
\centering
\includegraphics[width=1\columnwidth]{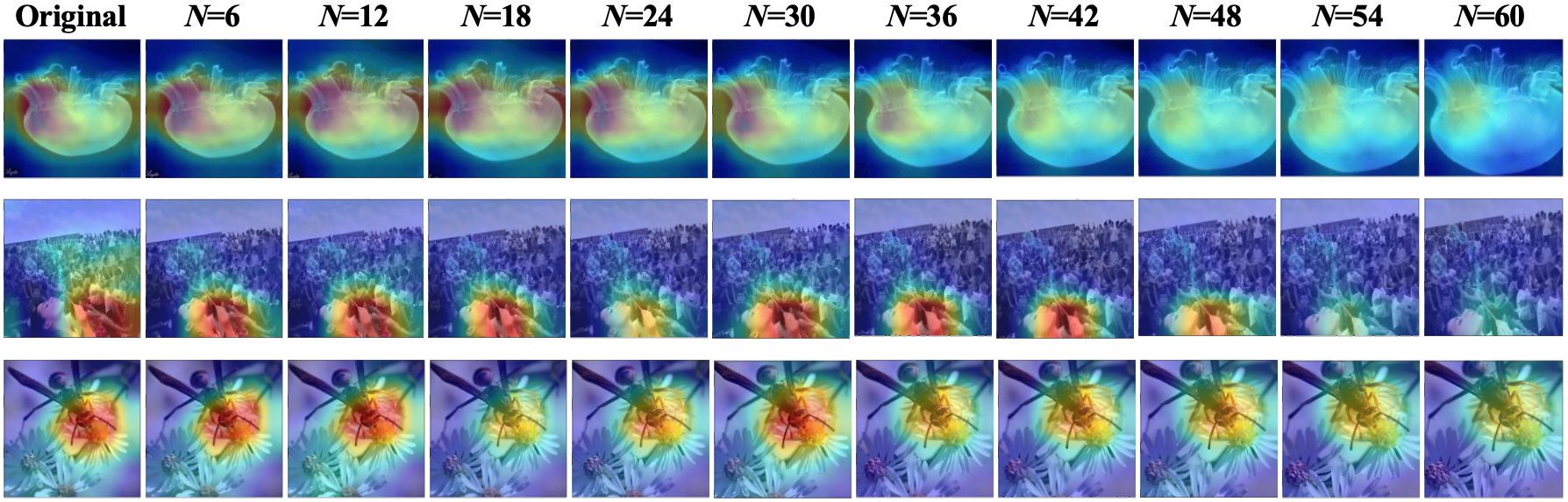} 
\caption{CAM for digital samples.}.
\label{figure7}
\end{figure}

We use CAM \cite{ref46} to demonstrate the model's attention display graph for adversarial samples in the digital environment. As can be seen from Figure \ref{figure7}, the model's attention is gradually weakened as the image is continuously enlarged. Here, we use clean samples with Jellyfish, Bubble and Bee, human observers are able to identify the correct category of the zoom-in digital samples, while classifiers are not. It means that the original model did not include images of the same category but different distances into the data set during training, leading the model to misclassifications of images at different distances.

\subsection{Defense of AdvZL}
In addition to describing the threat of the proposed AdvZL poses to advanced DNNs, we also try to propose a defense against AdvZL. Consistent with the idea of adversarial training, we construct a dataset for adversarial defense.

\begin{figure}[H]
\centering
\includegraphics[width=1\columnwidth]{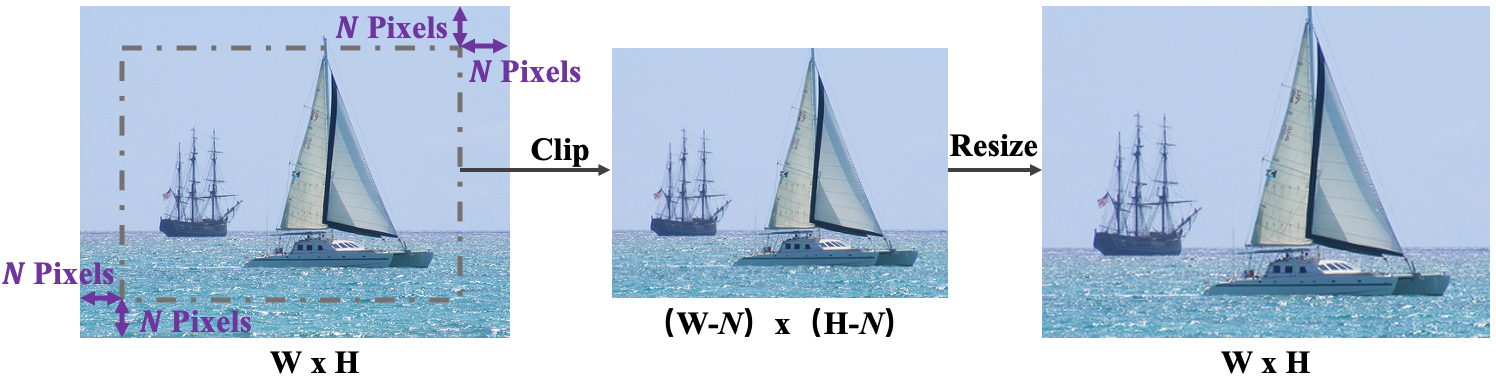} 
\caption{Zooming in image in the digital-setting.}.
\label{figure9}
\end{figure}

According to the above analysis, adversarial samples generated by AdvZL lead to discontinuous misclassification of models. Therefore, we build a scaled-up dataset derived from ImageNet called ImageNet-ZOOMIN (ImageNet-ZI). In which, 50 images are randomly selected from each category of ImageNet, resulting in 50, 000 clean samples. Then, each image is enlarged 10 different times to generate 500,000 adversarial samples. Among them, the approach of generating zoom-in samples is shown in Figure \ref{figure9}. Parameter $N$ ranges from 6 to 60 respectively, and the interval is 6. After the continuous magnification of the image, the semantic information of the zoom-in images is consistent with clean samples even only part of the target object on the images. Note that in the digital environment, we can only zoom in the image. Therefore, we only introduce the defense strategy against zoom in attack.

\begin{figure}[H]
\centering
\setlength{\belowcaptionskip}{0.2cm}
\includegraphics[width=1\columnwidth]{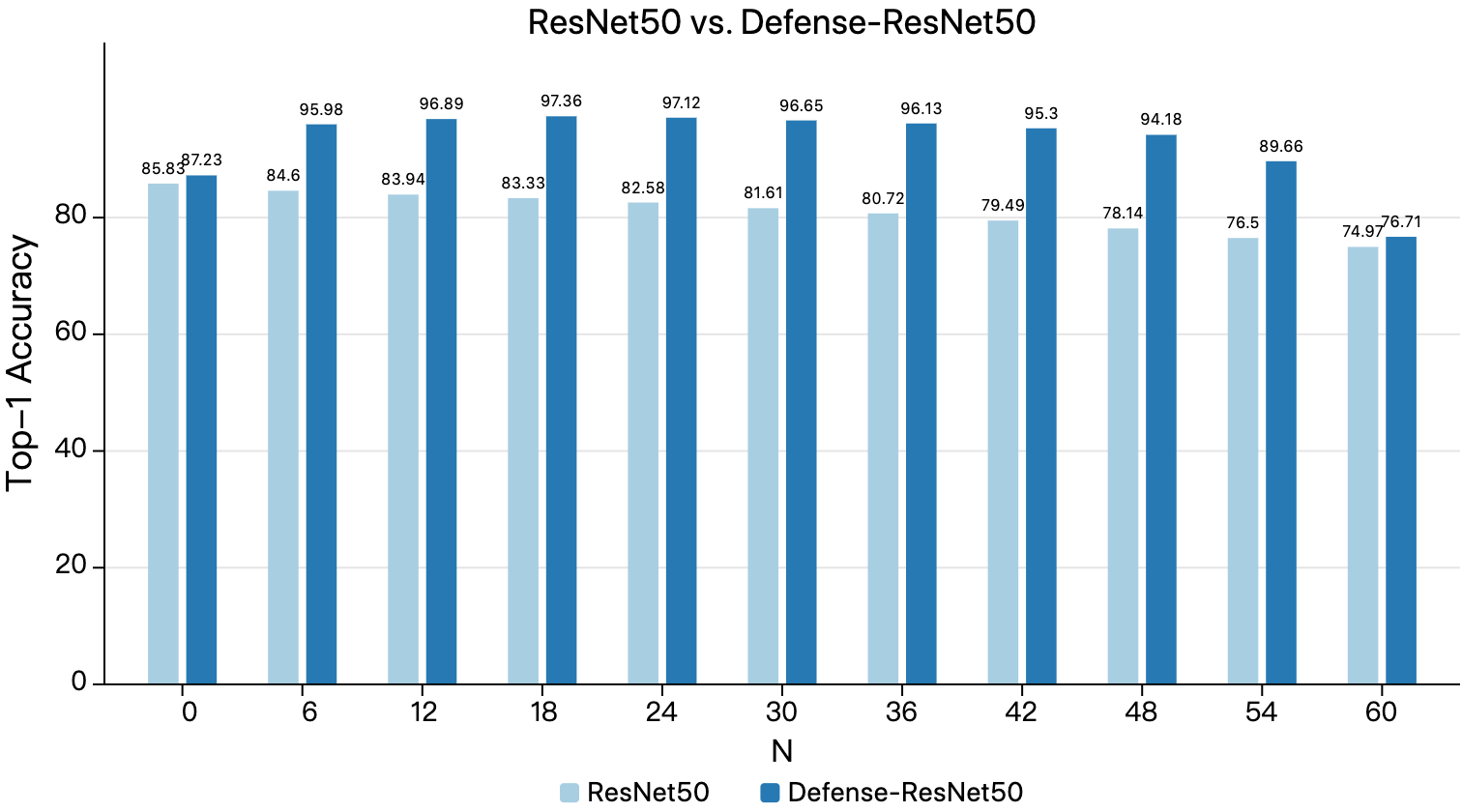} 
\caption{Performance of ResNet50 vs. Defense-ResNet50.}.
\label{figure10}
\end{figure}

We use torchvision to train the ResNet50 defense model. The model was optimized on 3 2080Ti GPUs by ADAM with initial learning rate 0.01. Here, we show a comparison of classification performance between the pretrained model of ResNet50 and adversarial trained ResNet50 (Defense-ResNet50). As can be seen from the experimental results in Figure \ref{figure10}, the model's robustness is improved to some extent through the strategy of adversarial training.

\subsection{Disadvantages of AdvZL}
We admit that in the digital environment, it’s currently unable to zoom out images through data processing techniques, which makes it hard to study zoom out attacks in the digital environment, which is a drawback of the proposed AdvZL. On the other hand, due to the limitation of conditions, we cannot deploy the zoom lens to the inside of the camera of the self-driving vehicle for physical attacks in real scenario, only use the camera of the mobile phone to simulate the attacks.

\section{Conclusion}
In this paper, we propose a camera-based physical attack, AdvZL, which generates adversarial samples without modifying target objects. Rigorous experimental design and comprehensive experimental results indicate the effective adversarial effect of the proposed AdvZL in both digital and physical environments. Our proposed method demonstrates the security threats to vision-based systems in the real world. If an attacker were to install an automatic zoom lens in the camera of an autonomous vehicle, the self-driving car will fail to recognize target objects, and it would be difficult for technicians to find out the cause of the accident. Our work offers some promising directions for some future physical attacks, manipulating camera to generate adversarial samples rather than manual deployment of physical perturbations. Our proposed AdvZL is very useful for studying the security threats of physical attacks to vision-based applications without modifying the target objects, which is a valuable addition to the current physical-world attacks.

In the future, we will continue to focus on improving the proposed AdvZL, for example, zoom out attacks in digital environments and deploy AdvZL to autonomous vehicles. In addition, we will continue to study physical attacks without adding physical perturbations, such as using lenses to achieve local distortion of images. We will also investigate the application of AdvZL to other areas such as object detection and segmentation. Finally, effective defenses against camera-based attacks will be a promising direction for the future.

\bibliography{aaai22}

\begin{thebibliography}{46}
\providecommand{\natexlab}[1]{#1}

\bibitem[{Athalye et~al.(2018)Athalye, Engstrom, Ilyas, and Kwok}]{ref31}
Athalye, A.; Engstrom, L.; Ilyas, A.; and Kwok, K. 2018.
\newblock Synthesizing Robust Adversarial Examples.
\newblock In Dy, J.~G.; and Krause, A., eds., \emph{Proceedings of the 35th
  International Conference on Machine Learning, {ICML} 2018,
  Stockholmsm{\"{a}}ssan, Stockholm, Sweden, July 10-15, 2018}, volume~80 of
  \emph{Proceedings of Machine Learning Research}, 284--293. {PMLR}.

\bibitem[{Brown et~al.(2017)Brown, Man{\'{e}}, Roy, Abadi, and Gilmer}]{ref29}
Brown, T.~B.; Man{\'{e}}, D.; Roy, A.; Abadi, M.; and Gilmer, J. 2017.
\newblock Adversarial Patch.
\newblock \emph{CoRR}, abs/1712.09665.

\bibitem[{Carlini and Wagner(2017{\natexlab{a}})}]{ref2}
Carlini, N.; and Wagner, D.~A. 2017{\natexlab{a}}.
\newblock Adversarial Examples Are Not Easily Detected: Bypassing Ten Detection
  Methods.
\newblock In Thuraisingham, B.; Biggio, B.; Freeman, D.~M.; Miller, B.; and
  Sinha, A., eds., \emph{Proceedings of the 10th {ACM} Workshop on Artificial
  Intelligence and Security, AISec@CCS 2017, Dallas, TX, USA, November 3,
  2017}, 3--14. {ACM}.

\bibitem[{Carlini and Wagner(2017{\natexlab{b}})}]{ref3}
Carlini, N.; and Wagner, D.~A. 2017{\natexlab{b}}.
\newblock Towards Evaluating the Robustness of Neural Networks.
\newblock In \emph{2017 {IEEE} Symposium on Security and Privacy, {SP} 2017,
  San Jose, CA, USA, May 22-26, 2017}, 39--57. {IEEE} Computer Society.

\bibitem[{Chen et~al.(2018{\natexlab{a}})Chen, Sharma, Zhang, Yi, and
  Hsieh}]{ref18}
Chen, P.; Sharma, Y.; Zhang, H.; Yi, J.; and Hsieh, C. 2018{\natexlab{a}}.
\newblock {EAD:} Elastic-Net Attacks to Deep Neural Networks via Adversarial
  Examples.
\newblock In McIlraith, S.~A.; and Weinberger, K.~Q., eds., \emph{Proceedings
  of the Thirty-Second {AAAI} Conference on Artificial Intelligence, (AAAI-18),
  the 30th innovative Applications of Artificial Intelligence (IAAI-18), and
  the 8th {AAAI} Symposium on Educational Advances in Artificial Intelligence
  (EAAI-18), New Orleans, Louisiana, USA, February 2-7, 2018}, 10--17. {AAAI}
  Press.

\bibitem[{Chen et~al.(2018{\natexlab{b}})Chen, Cornelius, Martin, and
  Chau}]{ref23}
Chen, S.; Cornelius, C.; Martin, J.; and Chau, D. H.~P. 2018{\natexlab{b}}.
\newblock ShapeShifter: Robust Physical Adversarial Attack on Faster {R-CNN}
  Object Detector.
\newblock In Berlingerio, M.; Bonchi, F.; G{\"{a}}rtner, T.; Hurley, N.; and
  Ifrim, G., eds., \emph{Machine Learning and Knowledge Discovery in Databases
  - European Conference, {ECML} {PKDD} 2018, Dublin, Ireland, September 10-14,
  2018, Proceedings, Part {I}}, volume 11051 of \emph{Lecture Notes in Computer
  Science}, 52--68. Springer.

\bibitem[{Deng et~al.(2009)Deng, Dong, Socher, Li, Li, and Fei{-}Fei}]{ref45}
Deng, J.; Dong, W.; Socher, R.; Li, L.; Li, K.; and Fei{-}Fei, L. 2009.
\newblock ImageNet: {A} large-scale hierarchical image database.
\newblock In \emph{2009 {IEEE} Computer Society Conference on Computer Vision
  and Pattern Recognition {(CVPR} 2009), 20-25 June 2009, Miami, Florida,
  {USA}}, 248--255. {IEEE} Computer Society.

\bibitem[{Dong et~al.(2018)Dong, Liao, Pang, Su, Zhu, Hu, and Li}]{ref4}
Dong, Y.; Liao, F.; Pang, T.; Su, H.; Zhu, J.; Hu, X.; and Li, J. 2018.
\newblock Boosting Adversarial Attacks With Momentum.
\newblock In \emph{2018 {IEEE} Conference on Computer Vision and Pattern
  Recognition, {CVPR} 2018, Salt Lake City, UT, USA, June 18-22, 2018},
  9185--9193. Computer Vision Foundation / {IEEE} Computer Society.

\bibitem[{Duan et~al.(2020)Duan, Ma, Wang, Bailey, Qin, and Yang}]{ref25}
Duan, R.; Ma, X.; Wang, Y.; Bailey, J.; Qin, A.~K.; and Yang, Y. 2020.
\newblock Adversarial Camouflage: Hiding Physical-World Attacks With Natural
  Styles.
\newblock In \emph{2020 {IEEE/CVF} Conference on Computer Vision and Pattern
  Recognition, {CVPR} 2020, Seattle, WA, USA, June 13-19, 2020}, 997--1005.
  Computer Vision Foundation / {IEEE}.

\bibitem[{Duan et~al.(2021)Duan, Mao, Qin, Chen, Ye, He, and Yang}]{ref35}
Duan, R.; Mao, X.; Qin, A.~K.; Chen, Y.; Ye, S.; He, Y.; and Yang, Y. 2021.
\newblock Adversarial Laser Beam: Effective Physical-World Attack to DNNs in a
  Blink.
\newblock In \emph{{IEEE} Conference on Computer Vision and Pattern
  Recognition, {CVPR} 2021, virtual, June 19-25, 2021}, 16062--16071. Computer
  Vision Foundation / {IEEE}.

\bibitem[{Eykholt et~al.(2018)Eykholt, Evtimov, Fernandes, Li, Rahmati, Xiao,
  Prakash, Kohno, and Song}]{ref24}
Eykholt, K.; Evtimov, I.; Fernandes, E.; Li, B.; Rahmati, A.; Xiao, C.;
  Prakash, A.; Kohno, T.; and Song, D. 2018.
\newblock Robust Physical-World Attacks on Deep Learning Visual Classification.
\newblock In \emph{2018 {IEEE} Conference on Computer Vision and Pattern
  Recognition, {CVPR} 2018, Salt Lake City, UT, USA, June 18-22, 2018},
  1625--1634. Computer Vision Foundation / {IEEE} Computer Society.

\bibitem[{Gnanasambandam, Sherman, and Chan(2021)}]{ref36}
Gnanasambandam, A.; Sherman, A.~M.; and Chan, S.~H. 2021.
\newblock Optical Adversarial Attack.
\newblock In \emph{{IEEE/CVF} International Conference on Computer Vision
  Workshops, {ICCVW} 2021, Montreal, BC, Canada, October 11-17, 2021}, 92--101.
  {IEEE}.

\bibitem[{Goodfellow, Shlens, and Szegedy(2015)}]{ref16}
Goodfellow, I.~J.; Shlens, J.; and Szegedy, C. 2015.
\newblock Explaining and Harnessing Adversarial Examples.
\newblock In Bengio, Y.; and LeCun, Y., eds., \emph{3rd International
  Conference on Learning Representations, {ICLR} 2015, San Diego, CA, USA, May
  7-9, 2015, Conference Track Proceedings}.

\bibitem[{He et~al.(2016)He, Zhang, Ren, and Sun}]{ref40}
He, K.; Zhang, X.; Ren, S.; and Sun, J. 2016.
\newblock Deep Residual Learning for Image Recognition.
\newblock In \emph{2016 {IEEE} Conference on Computer Vision and Pattern
  Recognition, {CVPR} 2016, Las Vegas, NV, USA, June 27-30, 2016}, 770--778.
  {IEEE} Computer Society.

\bibitem[{Hosseini and Poovendran(2018)}]{ref7}
Hosseini, H.; and Poovendran, R. 2018.
\newblock Semantic Adversarial Examples.
\newblock In \emph{2018 {IEEE} Conference on Computer Vision and Pattern
  Recognition Workshops, {CVPR} Workshops 2018, Salt Lake City, UT, USA, June
  18-22, 2018}, 1614--1619. Computer Vision Foundation / {IEEE} Computer
  Society.

\bibitem[{Huang et~al.(2017)Huang, Liu, van~der Maaten, and Weinberger}]{ref39}
Huang, G.; Liu, Z.; van~der Maaten, L.; and Weinberger, K.~Q. 2017.
\newblock Densely Connected Convolutional Networks.
\newblock In \emph{2017 {IEEE} Conference on Computer Vision and Pattern
  Recognition, {CVPR} 2017, Honolulu, HI, USA, July 21-26, 2017}, 2261--2269.
  {IEEE} Computer Society.

\bibitem[{Huang et~al.(2021)Huang, Liu, Yang, and Zhang}]{ref27}
Huang, S.; Liu, X.; Yang, X.; and Zhang, Z. 2021.
\newblock An improved ShapeShifter method of generating adversarial examples
  for physical attacks on stop signs against Faster R-CNNs.
\newblock \emph{Comput. Secur.}, 104: 102120.

\bibitem[{Krizhevsky, Sutskever, and Hinton(2012)}]{ref44}
Krizhevsky, A.; Sutskever, I.; and Hinton, G.~E. 2012.
\newblock ImageNet Classification with Deep Convolutional Neural Networks.
\newblock In Bartlett, P.~L.; Pereira, F. C.~N.; Burges, C. J.~C.; Bottou, L.;
  and Weinberger, K.~Q., eds., \emph{Advances in Neural Information Processing
  Systems 25: 26th Annual Conference on Neural Information Processing Systems
  2012. Proceedings of a meeting held December 3-6, 2012, Lake Tahoe, Nevada,
  United States}, 1106--1114.

\bibitem[{Kurakin, Goodfellow, and Bengio(2017)}]{ref22}
Kurakin, A.; Goodfellow, I.~J.; and Bengio, S. 2017.
\newblock Adversarial examples in the physical world.
\newblock In \emph{5th International Conference on Learning Representations,
  {ICLR} 2017, Toulon, France, April 24-26, 2017, Workshop Track Proceedings}.
  OpenReview.net.

\bibitem[{Li, Schmidt, and Kolter(2019)}]{ref38}
Li, J.; Schmidt, F.~R.; and Kolter, J.~Z. 2019.
\newblock Adversarial camera stickers: {A} physical camera-based attack on deep
  learning systems.
\newblock In Chaudhuri, K.; and Salakhutdinov, R., eds., \emph{Proceedings of
  the 36th International Conference on Machine Learning, {ICML} 2019, 9-15 June
  2019, Long Beach, California, {USA}}, volume~97 of \emph{Proceedings of
  Machine Learning Research}, 3896--3904. {PMLR}.

\bibitem[{Liu et~al.(2019)Liu, Tao, Li, Nowrouzezahrai, and Jacobson}]{ref15}
Liu, H.~D.; Tao, M.; Li, C.; Nowrouzezahrai, D.; and Jacobson, A. 2019.
\newblock Beyond Pixel Norm-Balls: Parametric Adversaries using an Analytically
  Differentiable Renderer.
\newblock In \emph{7th International Conference on Learning Representations,
  {ICLR} 2019, New Orleans, LA, USA, May 6-9, 2019}. OpenReview.net.

\bibitem[{Madry et~al.(2018)Madry, Makelov, Schmidt, Tsipras, and Vladu}]{ref5}
Madry, A.; Makelov, A.; Schmidt, L.; Tsipras, D.; and Vladu, A. 2018.
\newblock Towards Deep Learning Models Resistant to Adversarial Attacks.
\newblock In \emph{6th International Conference on Learning Representations,
  {ICLR} 2018, Vancouver, BC, Canada, April 30 - May 3, 2018, Conference Track
  Proceedings}. OpenReview.net.

\bibitem[{Moosavi{-}Dezfooli et~al.(2017)Moosavi{-}Dezfooli, Fawzi, Fawzi, and
  Frossard}]{ref21}
Moosavi{-}Dezfooli, S.; Fawzi, A.; Fawzi, O.; and Frossard, P. 2017.
\newblock Universal Adversarial Perturbations.
\newblock In \emph{2017 {IEEE} Conference on Computer Vision and Pattern
  Recognition, {CVPR} 2017, Honolulu, HI, USA, July 21-26, 2017}, 86--94.
  {IEEE} Computer Society.

\bibitem[{Moosavi{-}Dezfooli, Fawzi, and Frossard(2016)}]{ref17}
Moosavi{-}Dezfooli, S.; Fawzi, A.; and Frossard, P. 2016.
\newblock DeepFool: {A} Simple and Accurate Method to Fool Deep Neural
  Networks.
\newblock In \emph{2016 {IEEE} Conference on Computer Vision and Pattern
  Recognition, {CVPR} 2016, Las Vegas, NV, USA, June 27-30, 2016}, 2574--2582.
  {IEEE} Computer Society.

\bibitem[{Nguyen et~al.(2020)Nguyen, Arora, Wu, and Yang}]{ref33}
Nguyen, D.; Arora, S.~S.; Wu, Y.; and Yang, H. 2020.
\newblock Adversarial Light Projection Attacks on Face Recognition Systems: {A}
  Feasibility Study.
\newblock In \emph{2020 {IEEE/CVF} Conference on Computer Vision and Pattern
  Recognition, {CVPR} Workshops 2020, Seattle, WA, USA, June 14-19, 2020},
  3548--3556. Computer Vision Foundation / {IEEE}.

\bibitem[{Sandler et~al.(2018)Sandler, Howard, Zhu, Zhmoginov, and
  Chen}]{ref43}
Sandler, M.; Howard, A.~G.; Zhu, M.; Zhmoginov, A.; and Chen, L. 2018.
\newblock MobileNetV2: Inverted Residuals and Linear Bottlenecks.
\newblock In \emph{2018 {IEEE} Conference on Computer Vision and Pattern
  Recognition, {CVPR} 2018, Salt Lake City, UT, USA, June 18-22, 2018},
  4510--4520. Computer Vision Foundation / {IEEE} Computer Society.

\bibitem[{Shamsabadi, S{\'{a}}nchez{-}Matilla, and Cavallaro(2020)}]{ref8}
Shamsabadi, A.~S.; S{\'{a}}nchez{-}Matilla, R.; and Cavallaro, A. 2020.
\newblock ColorFool: Semantic Adversarial Colorization.
\newblock In \emph{2020 {IEEE/CVF} Conference on Computer Vision and Pattern
  Recognition, {CVPR} 2020, Seattle, WA, USA, June 13-19, 2020}, 1148--1157.
  Computer Vision Foundation / {IEEE}.

\bibitem[{Sharif et~al.(2016)Sharif, Bhagavatula, Bauer, and Reiter}]{ref30}
Sharif, M.; Bhagavatula, S.; Bauer, L.; and Reiter, M.~K. 2016.
\newblock Accessorize to a Crime: Real and Stealthy Attacks on State-of-the-Art
  Face Recognition.
\newblock In Weippl, E.~R.; Katzenbeisser, S.; Kruegel, C.; Myers, A.~C.; and
  Halevi, S., eds., \emph{Proceedings of the 2016 {ACM} {SIGSAC} Conference on
  Computer and Communications Security, Vienna, Austria, October 24-28, 2016},
  1528--1540. {ACM}.

\bibitem[{Shen et~al.(2019)Shen, Liao, Zhu, Xu, and Du}]{ref32}
Shen, M.; Liao, Z.; Zhu, L.; Xu, K.; and Du, X. 2019.
\newblock {VLA:} {A} Practical Visible Light-based Attack on Face Recognition
  Systems in Physical World.
\newblock \emph{Proc. {ACM} Interact. Mob. Wearable Ubiquitous Technol.}, 3(3):
  103:1--103:19.

\bibitem[{Simonyan and Zisserman(2015)}]{ref41}
Simonyan, K.; and Zisserman, A. 2015.
\newblock Very Deep Convolutional Networks for Large-Scale Image Recognition.
\newblock In Bengio, Y.; and LeCun, Y., eds., \emph{3rd International
  Conference on Learning Representations, {ICLR} 2015, San Diego, CA, USA, May
  7-9, 2015, Conference Track Proceedings}.

\bibitem[{Song et~al.(2018)Song, Eykholt, Evtimov, Fernandes, Li, Rahmati,
  Tram{\`{e}}r, Prakash, and Kohno}]{ref26}
Song, D.; Eykholt, K.; Evtimov, I.; Fernandes, E.; Li, B.; Rahmati, A.;
  Tram{\`{e}}r, F.; Prakash, A.; and Kohno, T. 2018.
\newblock Physical Adversarial Examples for Object Detectors.
\newblock In Rossow, C.; and Younan, Y., eds., \emph{12th {USENIX} Workshop on
  Offensive Technologies, {WOOT} 2018, Baltimore, MD, USA, August 13-14, 2018}.
  {USENIX} Association.

\bibitem[{Su, Vargas, and Sakurai(2019)}]{ref20}
Su, J.; Vargas, D.~V.; and Sakurai, K. 2019.
\newblock One Pixel Attack for Fooling Deep Neural Networks.
\newblock \emph{{IEEE} Trans. Evol. Comput.}, 23(5): 828--841.

\bibitem[{Szegedy et~al.(2015)Szegedy, Liu, Jia, Sermanet, Reed, Anguelov,
  Erhan, Vanhoucke, and Rabinovich}]{ref42}
Szegedy, C.; Liu, W.; Jia, Y.; Sermanet, P.; Reed, S.~E.; Anguelov, D.; Erhan,
  D.; Vanhoucke, V.; and Rabinovich, A. 2015.
\newblock Going deeper with convolutions.
\newblock In \emph{{IEEE} Conference on Computer Vision and Pattern
  Recognition, {CVPR} 2015, Boston, MA, USA, June 7-12, 2015}, 1--9. {IEEE}
  Computer Society.

\bibitem[{Szegedy et~al.(2014)Szegedy, Zaremba, Sutskever, Bruna, Erhan,
  Goodfellow, and Fergus}]{ref1}
Szegedy, C.; Zaremba, W.; Sutskever, I.; Bruna, J.; Erhan, D.; Goodfellow,
  I.~J.; and Fergus, R. 2014.
\newblock Intriguing properties of neural networks.
\newblock In Bengio, Y.; and LeCun, Y., eds., \emph{2nd International
  Conference on Learning Representations, {ICLR} 2014, Banff, AB, Canada, April
  14-16, 2014, Conference Track Proceedings}.

\bibitem[{Wang et~al.(2022)Wang, Jiang, Sun, Zhou, Gong, Zhang, Yao, and
  Chen}]{ref13}
Wang, D.; Jiang, T.; Sun, J.; Zhou, W.; Gong, Z.; Zhang, X.; Yao, W.; and Chen,
  X. 2022.
\newblock {FCA:} Learning a 3D Full-Coverage Vehicle Camouflage for Multi-View
  Physical Adversarial Attack.
\newblock In \emph{Thirty-Sixth {AAAI} Conference on Artificial Intelligence,
  {AAAI} 2022, Thirty-Fourth Conference on Innovative Applications of
  Artificial Intelligence, {IAAI} 2022, The Twelveth Symposium on Educational
  Advances in Artificial Intelligence, {EAAI} 2022 Virtual Event, February 22 -
  March 1, 2022}, 2414--2422. {AAAI} Press.

\bibitem[{Wang et~al.(2021)Wang, Liu, Yin, Liu, Tang, and Liu}]{ref11}
Wang, J.; Liu, A.; Yin, Z.; Liu, S.; Tang, S.; and Liu, X. 2021.
\newblock Dual Attention Suppression Attack: Generate Adversarial Camouflage in
  Physical World.
\newblock In \emph{{IEEE} Conference on Computer Vision and Pattern
  Recognition, {CVPR} 2021, virtual, June 19-25, 2021}, 8565--8574. Computer
  Vision Foundation / {IEEE}.

\bibitem[{Wiyatno and Xu(2018)}]{ref19}
Wiyatno, R.; and Xu, A. 2018.
\newblock Maximal Jacobian-based Saliency Map Attack.
\newblock \emph{CoRR}, abs/1808.07945.

\bibitem[{Wiyatno and Xu(2019)}]{ref10}
Wiyatno, R.; and Xu, A. 2019.
\newblock Physical Adversarial Textures That Fool Visual Object Tracking.
\newblock In \emph{2019 {IEEE/CVF} International Conference on Computer Vision,
  {ICCV} 2019, Seoul, Korea (South), October 27 - November 2, 2019},
  4821--4830. {IEEE}.

\bibitem[{Xie et~al.(2019)Xie, Zhang, Zhou, Bai, Wang, Ren, and Yuille}]{ref6}
Xie, C.; Zhang, Z.; Zhou, Y.; Bai, S.; Wang, J.; Ren, Z.; and Yuille, A.~L.
  2019.
\newblock Improving Transferability of Adversarial Examples With Input
  Diversity.
\newblock In \emph{{IEEE} Conference on Computer Vision and Pattern
  Recognition, {CVPR} 2019, Long Beach, CA, USA, June 16-20, 2019}, 2730--2739.
  Computer Vision Foundation / {IEEE}.

\bibitem[{Xu et~al.(2020)Xu, Zhang, Liu, Fan, Sun, Chen, Chen, Wang, and
  Lin}]{ref28}
Xu, K.; Zhang, G.; Liu, S.; Fan, Q.; Sun, M.; Chen, H.; Chen, P.; Wang, Y.; and
  Lin, X. 2020.
\newblock Adversarial T-Shirt! Evading Person Detectors in a Physical World.
\newblock In Vedaldi, A.; Bischof, H.; Brox, T.; and Frahm, J., eds.,
  \emph{Computer Vision - {ECCV} 2020 - 16th European Conference, Glasgow, UK,
  August 23-28, 2020, Proceedings, Part {V}}, volume 12350 of \emph{Lecture
  Notes in Computer Science}, 665--681. Springer.

\bibitem[{Zeng et~al.(2019)Zeng, Liu, Wang, Qiu, Xie, Tai, Tang, and
  Yuille}]{ref14}
Zeng, X.; Liu, C.; Wang, Y.; Qiu, W.; Xie, L.; Tai, Y.; Tang, C.; and Yuille,
  A.~L. 2019.
\newblock Adversarial Attacks Beyond the Image Space.
\newblock In \emph{{IEEE} Conference on Computer Vision and Pattern
  Recognition, {CVPR} 2019, Long Beach, CA, USA, June 16-20, 2019}, 4302--4311.
  Computer Vision Foundation / {IEEE}.

\bibitem[{Zhang et~al.(2019)Zhang, Foroosh, David, and Gong}]{ref12}
Zhang, Y.; Foroosh, H.; David, P.; and Gong, B. 2019.
\newblock {CAMOU:} Learning Physical Vehicle Camouflages to Adversarially
  Attack Detectors in the Wild.
\newblock In \emph{7th International Conference on Learning Representations,
  {ICLR} 2019, New Orleans, LA, USA, May 6-9, 2019}. OpenReview.net.

\bibitem[{Zhao, Liu, and Larson(2020)}]{ref9}
Zhao, Z.; Liu, Z.; and Larson, M.~A. 2020.
\newblock Towards Large Yet Imperceptible Adversarial Image Perturbations With
  Perceptual Color Distance.
\newblock In \emph{2020 {IEEE/CVF} Conference on Computer Vision and Pattern
  Recognition, {CVPR} 2020, Seattle, WA, USA, June 13-19, 2020}, 1036--1045.
  Computer Vision Foundation / {IEEE}.

\bibitem[{Zhong et~al.(2022)Zhong, Liu, Zhai, Jiang, and Ji}]{ref37}
Zhong, Y.; Liu, X.; Zhai, D.; Jiang, J.; and Ji, X. 2022.
\newblock Shadows can be Dangerous: Stealthy and Effective Physical-world
  Adversarial Attack by Natural Phenomenon.
\newblock In \emph{Proceedings of the IEEE/CVF Conference on Computer Vision
  and Pattern Recognition}, 15345--15354.

\bibitem[{Zhou et~al.(2016)Zhou, Khosla, Lapedriza, Oliva, and
  Torralba}]{ref46}
Zhou, B.; Khosla, A.; Lapedriza, {\`{A}}.; Oliva, A.; and Torralba, A. 2016.
\newblock Learning Deep Features for Discriminative Localization.
\newblock In \emph{2016 {IEEE} Conference on Computer Vision and Pattern
  Recognition, {CVPR} 2016, Las Vegas, NV, USA, June 27-30, 2016}, 2921--2929.
  {IEEE} Computer Society.

\bibitem[{Zhou et~al.(2018)Zhou, Tang, Wang, Han, Liu, and Zhang}]{ref34}
Zhou, Z.; Tang, D.; Wang, X.; Han, W.; Liu, X.; and Zhang, K. 2018.
\newblock Invisible Mask: Practical Attacks on Face Recognition with Infrared.
\newblock \emph{CoRR}, abs/1803.04683.

\end{thebibliography}

\end{document}